\documentclass[12pt,a4paper]{article}
\usepackage{float}
\usepackage{}
\usepackage{subfig}
\usepackage{amsmath}
\usepackage{amssymb}
\usepackage{graphicx}
\usepackage{hyperref}
\usepackage{csquotes}
\usepackage{multirow}
\usepackage[left=.45in,right=.45in,top=.6in,bottom=.6in,headheight=14.5pt]{geometry}
\usepackage{multirow}
\usepackage{pdflscape}
\usepackage[title]{appendix}
\usepackage[left=.45in,right=.45in,top=.6in,bottom=.6in,headheight=14.5pt]{geometry}
\usepackage{fmtcount} 
\usepackage{array,multirow}
\newcolumntype{C}[1]{>{\centering\arraybackslash}m{#1}}
\usepackage{epsfig}
\usepackage{float}
\usepackage{geometry}
\usepackage{amsmath}
\usepackage{cite}
\usepackage{orcidlink}
\usepackage{ctable}
\usepackage{comment}

\newenvironment{rcases}
{\left.\begin{aligned}}
{\end{aligned}\right\rbrace}

\usepackage{hyperref}
\newgeometry{vmargin={10mm,20mm},hmargin={19mm,19mm}}

\begin{document}


\title{\textcolor{blue}{Reconciling TM$_2$ Mixing with LMA and Dark-LMA Data based on Minimal Corrections from Charged-Lepton Sector}}

\author{Ayush Kumar Singh\orcidlink{0009-0002-2468-7225}\thanks{singhayush138421@gmail.com }, Tapender\orcidlink{0009-0006-4109-528X}\thanks{tapenderphy@gmail.com (corresponding author)}, Labh Singh\orcidlink{0000-0001-7704-726X}\thanks{sainilabh5@gmail.com}  and Surender Verma\orcidlink{0000-0002-5671-5369}\thanks{s\_7verma@hpcu.ac.in}}

\date{%
Department of Physics and Astronomical Science\\
Central University of Himachal Pradesh\\
Dharamshala, India 176215
}
\maketitle
\begin{abstract}
\noindent Motivated by the increasing precision of neutrino oscillation data, we study the corrections to the TM$_2$ neutrino mixing framework, emanating from $(1,2)$ sector of the charged lepton, for both the standard LMA and dark-LMA solutions. We employ the Wolfenstein parameterization of the charged-lepton mixing matrix, characterized by two additional parameters $(\lambda,\delta)$, which effectively reconciles the TM$_2$ neutrino-mixing predictions with current oscillation data. For the LMA solution, the allowed ranges are $0.1 \lesssim \lambda \lesssim 0.33$ and $\delta \in (20^\circ\!-\!90^\circ)\oplus(270^\circ\!-\!340^\circ)$, while the dark-LMA case requires $\lambda>0.24$ and $125^\circ<\delta<235^\circ$. Interestingly, for LMA case, the upper bound $\lambda \le 0.33$ is found to be dictated by the atmospheric mixing angle $\theta_{23}$. The model predicts sizeable CP violation, with $|J_{CP}|$ reaching values as large as $0.13$. We, also, analyze the effective Majorana mass parameter $m_{ee}$ relevant for neutrinoless double beta decay. The inverted hierarchy region lies within the sensitivity of future experiments for both solutions, whereas only part of the normal hierarchy region can be tested.
\end{abstract}
\section{Introduction}
\noindent The non-zero neutrino mass unavoidably require that the structure of neutrino mixing matrix must be non-trivial. In general, neutrino mixing matrix contains four parameters \textit{viz.,} three mixing angles and one Dirac CP phase which describe the extent to which flavor and mass eigenstates are mixed. The neutrino oscillation experiments have been extremely successful in measuring three mixing angles with unprecedented precision. Thus, except CP phase, the dominant structure of the mixing matrix is known. 
From the theoretical point of view, the origin of observed mixing structure is still unknown which lead to proposition of several phenomenological ansatze some of which have, already been excluded by current experimental data (example is Tri-bimaximal mixing (TBM)~\cite{Harrison:2002er,Harrison:2002kp,Xing:2002sw}). Another, interesting ansatze for the mixing matrix is trimaximal mixing structure (TM$_2$) which results in magic structure of the neutrino mass matrix~\cite{Albright:2008rp,Albright:2010ap,Kumar:2010qz,Lam:2006wy,Tapender:2024ktc,King:2011zj,Zhao:2021dwc,He:2011gb,King:2015aea,Singh:2022nmk,Verma:2019uiu,Hyodo:2022xyn,Channey:2018cfj,Gautam:2016qyw}. Further, TM$_2$ is phenomenologically appealing, as it accommodates a non-zero reactor angle and permits leptonic CP violation measurable in neutrino oscillation experiments.    

\noindent Nevertheless, pure TM$_2$ predicts a solar mixing angle $\sin^2\theta_{12}$ that is typically larger than current
global-fit values~\cite{Esteban:2024eli}, leading to the so-called solar tension~\cite{Chen:2023hmn}. Resolving this anomaly becomes increasingly important in view of the increased precision with which current and upcoming neutrino oscillation experiments are reporting their measurements. Thus, mixing frameworks, in general, must be in perfect alignment with these observations and it is timely to investigate for possible deviation to the TM$_2$ mixing structure at the subdominant level. Various approaches have been explored in this direction, for example, Ref.~\cite{Fodroci:2025qgq} introduced deviations directly within the TM$_2$ matrix \textit{via} a real perturbation parameter.

\noindent Further, the level of sensitivity of neutrino experiments, as discussed above, not only sharpens tests of standard mixing scenarios but, also, provides a powerful probe of alternative solutions to solar neutrino oscillations, such as the dark-LMA solution~\cite{Farzan:2015doa,Choubey:2019osj,Ghosh:2020fes,Denton:2022nol,Denton:2018xmq,Coloma:2017egw,Deepthi:2019ljo,Kumar:2025kgd,Singh:2022ijf}. The dark-LMA scenario may arise in the presence of non-standard neutrino interactions and corresponds to a distinct region of parameter space that can mimic the standard LMA solution in oscillation experiments. Although recent data including coherent elastic neutrino–nucleus scattering (CE$\nu$NS) measurements~\cite{Chaves:2021pey} and reactor experiments~\cite{Denton:2022nol} have significantly constrained this degeneracy, it is not completely excluded. In particular, viable parameter space remains in scenarios with non-universal couplings or in the $\nu_\mu$ and $\nu_\tau$ sectors, leaving open the possibility of dark-LMA like solutions. The possible existence of this solution, thus, warrants scrutiny as the allowable deviation in the TM$_2$ mixing paradigm may open up new regions of parameter space depending on the parameters that characterize the deviation. Therefore, it is imperative to study the phenomenology of this scenario under the proposed deviation in TM$_2$ discussed below. 

\noindent In general, the charged-lepton mass matrix need not be diagonal as assumed in earlier analyses~\cite{Fodroci:2025qgq,Hyodo:2023sku,Chen:2023hmn}. In this work, we consider minimal deviations arising from the charged-lepton sector by employing a Wolfenstein-like parametrization of the charged-lepton mixing matrix~\cite{He:2009jm,Ma:2012zma,Liu:2014gla}. The approach is minimal in the sense that we restrict ourselves to deviations originating solely from the (1,2) sector of the charged-lepton mixing. These corrections offer an economical way to deviate from the leading-order trimaximal structure while yielding mixing angles in consonance with current data. We analyze this framework in both the standard LMA and the dark-LMA regimes.  

\noindent The remaining structure of the paper is as follows. In Sec.~2, we present the formalism leading to charge lepton corrections to the TM$_2$ mixing matrix. In Sec.~3, we provide the numerical analysis and discuss the results for both the LMA and dark-LMA solutions. Finally, we summarize our findings in Sec.~4.

\section{Formalism: Corrections to TM$_2$ mixing from charged lepton sector}
 As discussed in the introduction the neutrino mixing matrix considered in this work is based on the trimaximal mixing scheme of type TM$_2$, in which the second column of the mixing matrix remains identical to that of the tri-bi maximal (TBM) matrix. The TM$_2$ mixing matrix is parameterized as
\begin{equation}
U_{TM_2}=\left(
\begin{array}{ccc}
 \sqrt{\frac{2}{3}} \cos \theta  & \frac{1}{\sqrt{3}} & \sqrt{\frac{2}{3}} \sin \theta  \\
 -\frac{\cos \theta }{\sqrt{6}}+\frac{e^{-i \phi } \sin \theta }{\sqrt{2}} & \frac{1}{\sqrt{3}} & -\frac{\sin \theta}{\sqrt{6}}-\frac{e^{-i \phi } \cos \theta }{\sqrt{2}} \\
 -\frac{\cos \theta }{\sqrt{6}}-\frac{e^{-i \phi } \sin \theta }{\sqrt{2}} & \frac{1}{\sqrt{3}} & -\frac{\sin \theta}{\sqrt{6}}+\frac{e^{-i \phi } \cos \theta }{\sqrt{2}} \\
\end{array}
\right),
\label{eq:1}
\end{equation}
where $\theta$ and $\phi$ are real free parameters. 
However, pure TM$_2$ mixing predicts a solar mixing angle $\theta_{12}$ that is larger than the experimentally observed value, leading to the so-called {solar tension}.
To address this issue, we consider corrections arising from the charged lepton sector.
In analogy with the Wolfenstein parametrization of the quark mixing matrix\cite{Xing:1994rj}, a Wolfenstein-like parametrization in the charged lepton sector assumes that the charged lepton diagonalization matrix can be expanded in powers of a small dimensionless  parameter $\lambda$\cite{Xing:2002az}. In this framework, the charged lepton diagonalization matrix is parametrized as a unitary matrix expanded perturbatively in powers of the small real parameter $\lambda$, retaining terms up to $\lambda^{2}$ as 
\begin{equation}
U_{lep}=\left(
\begin{array}{ccc}
 1-\frac{\lambda ^2}{2} & \lambda  e^{i \delta } & 0 \\
 -\lambda  e^{-i \delta } & 1-\frac{\lambda ^2}{2} & 0 \\
 0 & 0 & 1 \\
\end{array}
\right),
\label{eq:2}
\end{equation}
where $\lambda$ and $\delta$ are real free parameters.
The physical lepton mixing matrix, known as the Pontecorvo–Maki–Nakagawa–Sakata (PMNS) matrix, is obtained as
\begin{eqnarray}
U_{PMNS} = U_{lep}^\dagger U_{TM_2} =
\left(
\begin{array}{ccc}
 U_{11} & U_{12} & U_{13} \\
 U_{21} & U_{22} & U_{23} \\
 U_{31} & U_{32} & U_{33} \\
\end{array}
\right),
\label{eq:3}
\end{eqnarray}
where $U_{ij}$ denote the elements of the PMNS matrix. This form of the PMNS matrix is a model independent unitary matrix. This framework allows us to study deviations from exact TM$_2$ mixing and introduces an additional phase, $\delta$, in U$_{PMNS}$ matrix which has important implications for CP violation.
However, it should be noted that these deviations do not affect the neutrino mass matrix, which remains a magic matrix. The elements of PMNS matrix  can be expressed as follows
\begin{equation}\label{eq:4}
\begin{rcases}
    U_{11}&=\sqrt{\frac{2}{3}} \left(1-\frac{\lambda ^2}{2}\right) \cos \theta-\lambda  e^{i \delta } \left(-\frac{\cos \theta
   }{\sqrt{6}}+\frac{e^{-i \phi } \sin \theta }{\sqrt{2}}\right),\\
     U_{12}&= \frac{1-\frac{\lambda ^2}{2}}{\sqrt{3}}-\frac{\lambda  e^{i \delta
   }}{\sqrt{3}},\\   
     U_{13}&=\sqrt{\frac{2}{3}} \left(1-\frac{\lambda ^2}{2}\right) \sin \theta -\lambda  e^{i \delta }
   \left(-\frac{\sin \theta }{\sqrt{6}}-\frac{e^{-i \phi } \cos \theta }{\sqrt{2}}\right), \\
     U_{21}&= \sqrt{\frac{2}{3}} \lambda  e^{-i \delta } \cos \theta +\left(1-\frac{\lambda ^2}{2}\right) \left(-\frac{\cos \theta
   }{\sqrt{6}}+\frac{e^{-i \phi } \sin \theta }{\sqrt{2}}\right),\\
     U_{22}&= \frac{1-\frac{\lambda ^2}{2}}{\sqrt{3}}+\frac{\lambda  e^{-i \delta
  }}{\sqrt{3}},\\ 
     U_{23}&= \sqrt{\frac{2}{3}} \lambda  e^{-i \delta } \sin \theta +\left(1-\frac{\lambda ^2}{2}\right)
   \left(-\frac{\sin \theta }{\sqrt{6}}-\frac{e^{-i \phi } \cos \theta }{\sqrt{2}}\right),\\ 
     U_{31}&=-\frac{\cos \theta }{\sqrt{6}}-\frac{e^{-i \phi } \sin \theta }{\sqrt{2}},\\
     U_{32}&=  \frac{1}{\sqrt{3}},\\
     U_{33}&=  -\frac{\sin \theta
   }{\sqrt{6}}+\frac{e^{-i \phi } \cos \theta}{\sqrt{2}}.
\end{rcases}
\end{equation}
\noindent The neutrino mixing angles can be determined from the PMNS matrix using the following relations
\begin{eqnarray}\label{angle}
\sin^{2}{\theta_{13}}=\left|U_{13}\right|^{2}, \hspace{0.5cm} \sin^{2}{\theta_{23}}=\dfrac{\left|U_{23}\right|^{2}}{1-\left|U_{13}\right|^{2}}, \hspace{0.5cm} \sin^{2}{\theta_{12}}=\dfrac{\left|U_{12}\right|^{2}}{1-\left|U_{13}\right|^{2}}.
\label{eqn16}
\end{eqnarray}
The degree of CP-violation quantified using the Jarlskog invariant is defined as
\begin{equation}
J_{\text{CP}} = \text{Im}\left[U_{11}U_{22}U^{*}_{12}U^{*}_{21}\right].
\end{equation}
The CP-violating effects are proportional to the Jarlskog invariant, which is geometrically associated with twice the area of the unitarity triangles \cite{Ellis:2020ehi}.  A non-zero value of $J_{\text{CP}}$ is a necessary and sufficient condition for CP violation. 
\noindent The detection of neutrinoless double beta decay ($0\nu\beta\beta$) would provide direct evidence for the violation of total lepton number and would demonstrate the Majorana nature of neutrinos.
The effective Majorana mass is defined as 
\begin{equation}
m_{ee}
=
\left|
U_{11}^2\, m_1
+
U_{12}^2\, m_2
+
U_{13}^2\, m_3
\right|, 
\end{equation}
where $m_i \,(i=1,2,3)$ represent the masses of the three light neutrino mass eigenstates, and $U_{1i}$ correspond to the elements of the first row of the PMNS mixing matrix \cite{Bilenky:2014uka}. In the next section, we examine the phenomenological implications of the correction to TM$_2$ mixing and will obtain prediction on these observables.

\section{Numerical analysis and Discussion}
To study the phenomenological viability of the proposed lepton mixing structure, we perform a numerical analysis by scanning the relevant parameter space of the model. The numerical study is carried out using a Monte Carlo statistical sampling approach wherein  $10^{7}$ random parameter points are generated and the predictions are tested against the latest experimental constraints from neutrino oscillation data. The free parameters of the model are varied within physically motivated intervals and sampled using uniform random distributions over the specified ranges. The ranges of the free parameters used in the numerical analysis are summarized in Table~\ref{tab:free_parameters}. 
In order to obtain the neutrino masses, we use the following relations for the normal hierarchy (NH) and inverted hierarchy (IH)
\begin{align}
\text{NH:} \qquad
m_2 &= \sqrt{m_{\text{lightest}}^2 + \Delta m_{21}^2}, &
m_3 &= \sqrt{m_{\text{lightest}}^2 + \Delta m_{31}^2}, \\
\text{IH:} \qquad
m_1 &= \sqrt{m_{\text{lightest}}^2 + |\Delta m_{31}^2|}, &
m_2 &= \sqrt{m_{\text{lightest}}^2 + |\Delta m_{31}^2| + \Delta m_{21}^2}.
\end{align}
The experimentally measured neutrino mass-squared differences are generated using Gaussian distributions centered around their respective best-fit values shown in Table~\ref{tab:nudata}. 
The neutrino mixing angles are constrained within their $3\sigma$ ranges using the data presented in Table~\ref{tab:nudata}.
In addition, we have, also, used cosmological bound on sum of neutrino masses, $\sum_i m_i<0.12$ eV\cite{Planck:2018vyg} to obtain allowed parameter space.  

\begin{figure}[t!]
    \centering
    \includegraphics[width=0.9\linewidth]{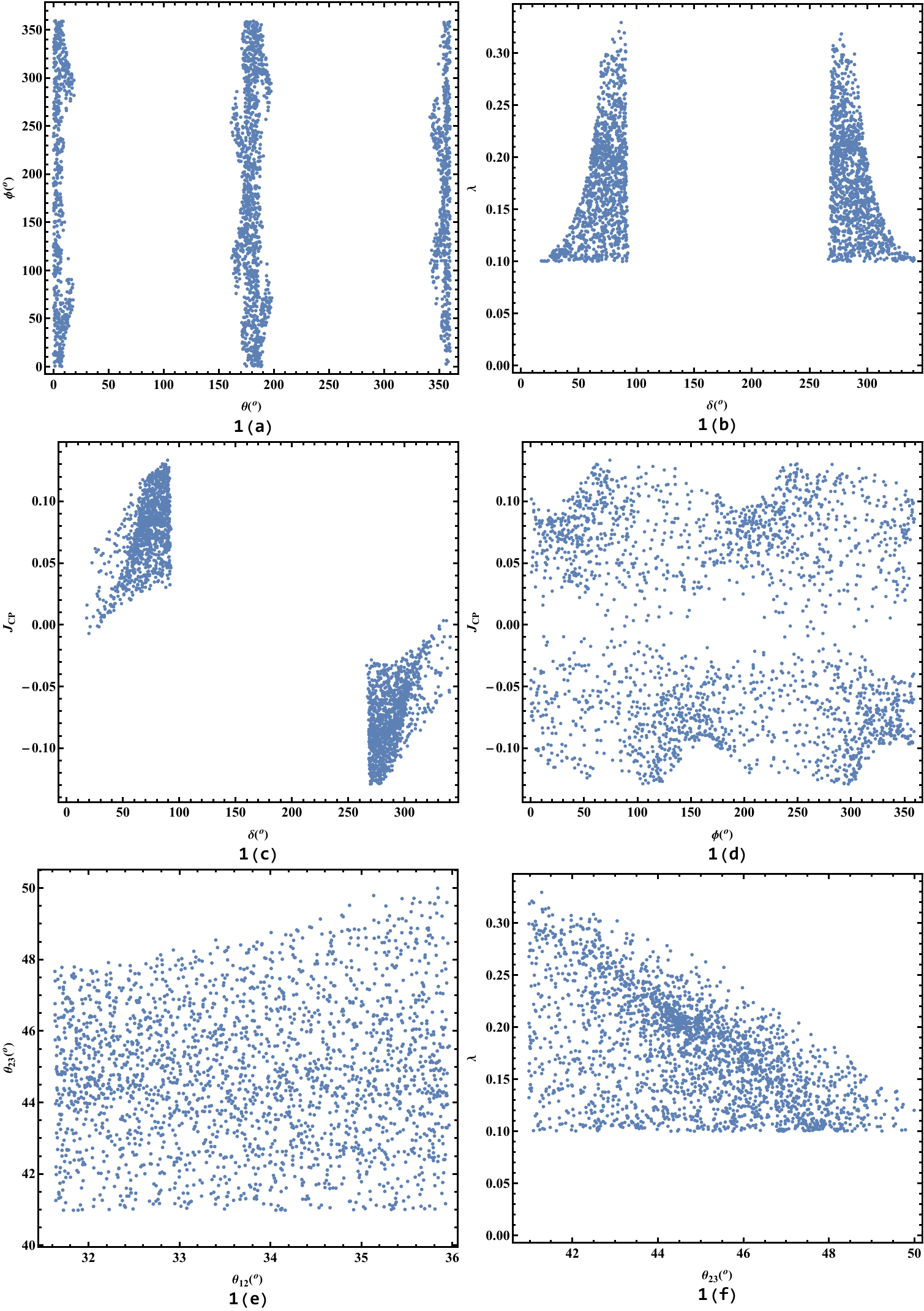}
    \caption{The correlation between different parameters for the NH-LMA solution is shown.}
    \label{fig1}
\end{figure}

\begin{figure}[t!]
    \centering
    \includegraphics[width=0.9\linewidth]{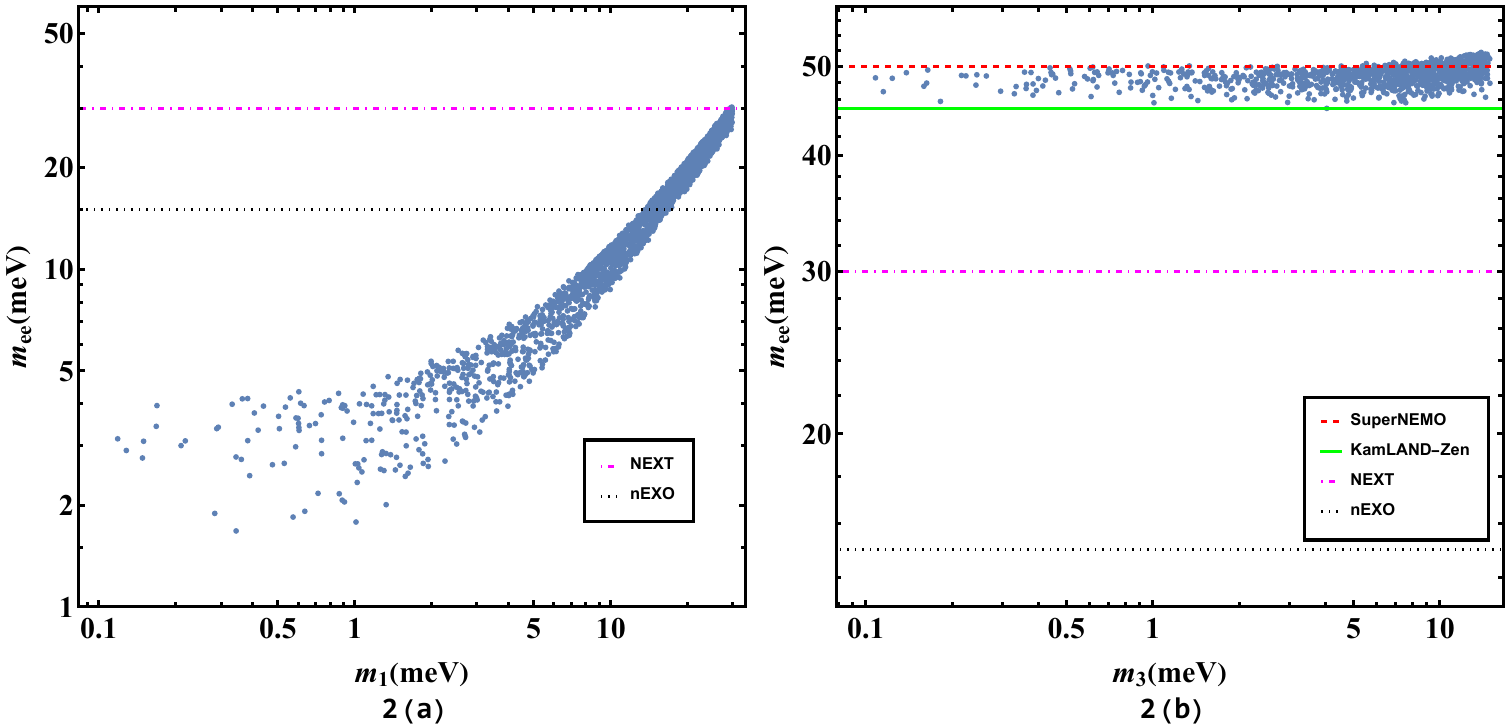}
    \caption{The correlation between the effective Majorana mass $m_{ee}$ is shown as a function of the lightest neutrino mass, $m_1$ for the NH-LMA solution (Fig. \ref{fig2}(a)) and $m_3$ for the IH-LMA solution (Fig. \ref{fig2}(b)). The horizontal lines represent the experimental sensitivities of neutrinoless double beta decay ($0\nu\beta\beta$) experiments.}
    \label{fig2}
\end{figure}

\begin{figure}[t!]
    \centering
    \includegraphics[width=0.9\linewidth]{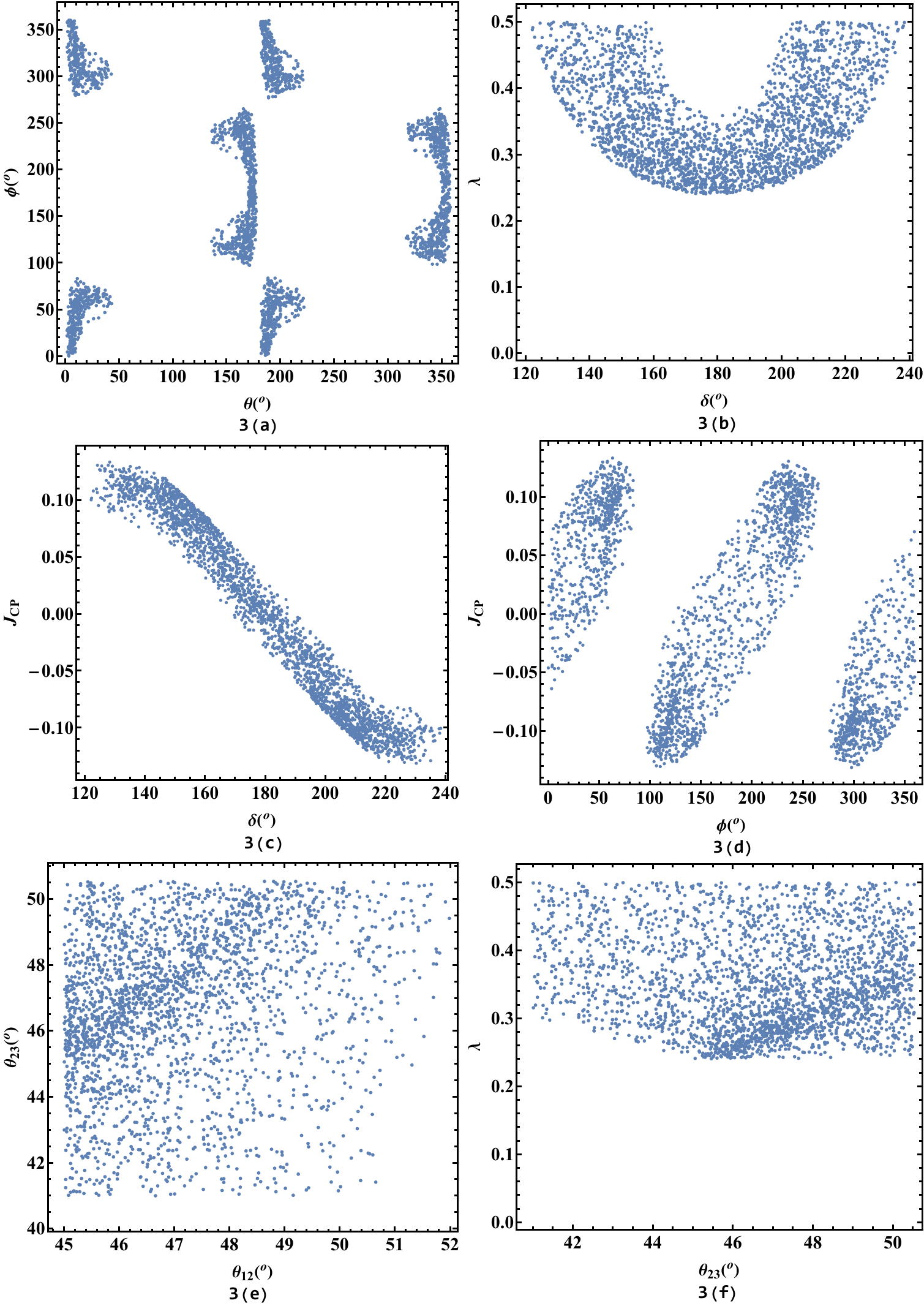}
    \caption{The correlation between different parameters for NH-dark LMA solution.}
    \label{fig3}
\end{figure}

\begin{figure}[t!]
    \centering
    \includegraphics[width=0.9\linewidth]{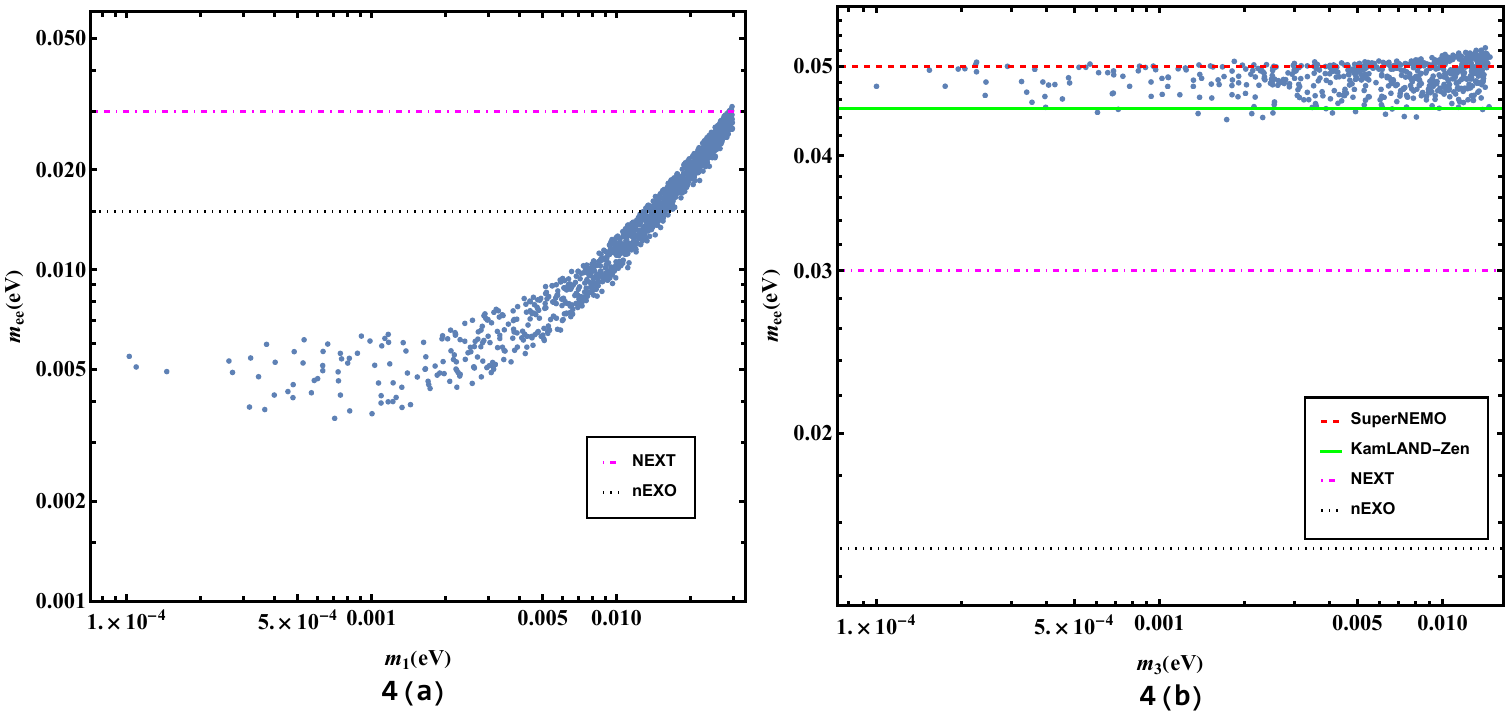}
    \caption{The correlation between the effective Majorana mass $m_{ee}$ is shown as a function of the lightest neutrino mass, $m_1$ for NH-dark LMA solutions (Fig. \ref{fig4}(a)) and $m_3$ for IH-dark LMA solutions (Fig. \ref{fig4}(b)). The horizontal lines represent the experimental sensitivities of neutrinoless double beta decay ($0\nu\beta\beta$) experiments.}
    \label{fig4}
\end{figure}

\begin{table}[t]
\centering

\begin{tabular}{|c|c|}
\hline
Parameter & Range \\ \hline
$\lambda$ & $0.1 \leq \lambda \leq 0.5$ \\ \hline
$\theta$ & $0^\circ \leq \theta \leq 360^\circ$ \\ \hline
$\phi$ & $0^\circ \leq \phi \leq 360^\circ$ \\ \hline
$\delta$ & $0^\circ \leq \delta \leq 360^\circ$ \\ \hline
$m_{\text{lightest}}$ & $10^{-4}\text{eV} \leq m_{\text{lightest}} \leq 0.1\ \text{eV}$ \\ \hline
\end{tabular}
\caption{Ranges of the free parameters used in the numerical analysis.}
\label{tab:free_parameters}
\end{table}

\begin{table}[t]
\small
    \centering
    \renewcommand{\arraystretch}{1} 
    \begin{tabular}{|c|cc|cc|} 
        \hline
        Parameter & best-fit$\pm1\sigma$ range & best-fit$\pm1\sigma$ range & $3\sigma$ range (NH) & $3\sigma$ range (IH) \\
 & (NH)& (IH)& &\\
        \hline
        $\sin^2\theta_{12}$ & $0.308^{+0.012}_{-0.011}$ & $0.308^{+0.012}_{-0.011}$ & $0.275 - 0.345$ & $0.275 - 0.345$ \\
        $\sin^2\theta_{23}$ & $0.470^{+0.017}_{-0.013}$ & $0.562^{+0.012}_{-0.015}$ & $0.435 - 0.585$ & $0.410 - 0.623$ \\
        $\sin^2\theta_{13}$ & $0.02215^{+0.00056}_{-0.00058}$ & $0.02224^{+0.00056}_{-0.00057}$ & $0.02023 - 0.02388$ & $0.02053 - 0.02397$ \\
        $\frac{\Delta m^2_{31} } {10^{-3} \text{eV}^2}$& $2.513^{+0.021}_{-0.019}$ & $-2.510^{+0.024}_{-0.025}$ & $2.463 - 2.606$ & $-2.584 - -2.438$ \\
        $\frac{\Delta m^2_{21}} {10^{-5} \text{eV}^2}$& $7.49^{+0.19}_{-0.19}$ & $7.49^{+0.19}_{-0.19}$ & $6.92 - 8.05$ & $6.92 - 8.05$ \\
        \hline
    \end{tabular}
    \caption{The experimental neutrino oscillation data used in the numerical analysis\cite{Esteban:2024eli}.}
    \label{tab:nudata}
\end{table}

\noindent We have studied both solutions, LMA and dark-LMA, under the minimal correction received from (1, 2) sector of the charged lepton sector. 
There are four free parameters,  $\theta$ and $\phi$, arising from the trimaximal nature of neutrino mixing matrix (TM$_2$), and two additional parameters, $\lambda$ and $\delta$, originating from the charged lepton mixing matrix. Fig.~\ref{fig1}  shows the correlations among these parameters for the normal hierarchy LMA solutions. 
As evident from Figs.~\ref{fig1}(a), phase $\phi$ can take any value in the range $0^\circ$--$360^\circ$
, however, $\theta$ can only be around $0^\circ, 180^\circ$ and $360^\circ$. Figs.~\ref{fig1}(b) explains the correlation between the parameters quantifying the correction to trimaximal mixing. While phase $\delta$ is constrained to the ranges $(20^\circ-90^\circ)\oplus (270^\circ-340^\circ)$, $\lambda$ is restricted to $(0.1-0.33)$ compared to its input range $(0.1-0.5)$ (Table~\ref{tab:free_parameters}). Interestingly, the upper bound $\lambda \le 0.33$ is predominantly dictated by the atmospheric mixing angle $\theta_{23}$, as evidenced by the correlation between $\theta_{23}$ and $\lambda$ illustrated in Fig.~\ref{fig1}(f). As one moves from higher to lower $\theta_{23}$ values in its predicted 3$\sigma$ range the upper bound on $\lambda$ increases, however, $\lambda>0.33$ will push $\theta_{23}$ below 3$\sigma$ lower limit and, thus, is disallowed. The current best-fit value of $\theta_{23}=48.5^\circ$ is consistent with the upper bound $\lambda\le0.17$.

\noindent In Figs.~\ref{fig1}(c,d), we have shown the  Jarlskog invariant parameter ($J_{CP}$) correlation with $\delta$ and $\phi$. The density of the allowed points clearly indicate CP violating nature of the correction, in fact, maximal $|J_{CP}|$ lies around $0.13$, with the phases $\phi$  and $\delta$ allowed in the regions stated earlier. In Fig.~\ref{fig1}(e), $\theta_{12}-\theta_{23}$ correlation is shown. An interesting feature of this plot is that while the lower limit of $\theta_{23}$ remains unaffected, upper 3$\sigma$ limit increases as one moves from lower to higher $\theta_{12}$ values in its allowed physical range. The plots for IH have similar correlative features and thus, are not presented here.

\noindent Furthermore, the correlation between the effective neutrino mass $m_{ee}$ and the lightest neutrino mass, for NH and IH, is shown in Fig.~\ref{fig2}(a,b) respectively. The horizontal lines in this figure represent the experimental sensitivities of various neutrinoless double beta decay ($0\nu\beta\beta$) experiments:  NEXT~\cite{NEXT:2013wsz,NEXT:2009vsd}, nEXO~\cite{Licciardi:2017oqg}, KamLAND-Zen~\cite{KamLAND-Zen:2016pfg} and SuperNEMO~\cite{Barabash:2011row}. For the NH case, only a portion of the allowed parameter space can be probed by the NEXT and nEXO $0\nu\beta\beta$ experiments. The remaining allowed region lies beyond the projected sensitivity of future $0\nu\beta\beta$ decay experiments, making the NH scenario difficult to probe completely. In contrast, for the IH case, the predicted parameter range falls within the sensitivity of current and upcoming experiments and can therefore be fully probed.

\noindent In general, the correction proposed in the present work, to make TM$_2$ predicted neutrino mixing parameters in consonance with corresponding experimental range(s), may induce larger values of mixing angles. Of particular interest is the solar sector mixing angle, $\theta_{12}$. As discussed earlier, the presence of NSI may result in the value of $\sin^2\theta_{12}\approx 0.7$ (dark-LMA solution), which is about  2.3 times the value corresponding to LMA solution (Table~\ref{tab:nudata}). It is, therefore, natural to examine the phenomenological implications arising from charged-lepton corrections in the (1,2) sector for assessing the viability of the dark-LMA solution. The numerical analysis is identical to that employed in obtaining Fig.~\ref{fig1}, except that $\sin^2\theta_{12}$ is now required to lie within the range $(0.5–0.8)$ in order to obtain a allowed parameter space of the model. 
In Fig.~\ref{fig3} we have shown the correlation plots for this solution. The main results of this analysis are: (i) for $\theta \simeq (0^\circ\!-\!40^\circ)$ or $(180^\circ\!-\!220^\circ)$: $\phi \in (0^\circ\!-\!90^\circ)\oplus(280^\circ\!-\!360^\circ)$, and for $\theta \simeq (140^\circ\!-\!180^\circ)$ or $(320^\circ\!-\!360^\circ)$: $\phi \in (90^\circ\!-\!280^\circ)$, (ii) CP phase $\delta$ is restricted to the range $(125^\circ<\delta<235^\circ)$ while there exist a lower bound on $\lambda>0.24$ which initially is uniformly varied between ($0.1-0.5$), (iii) 
above $\lambda=0.24$, $\theta_{23}$ can lie above or below maximality, however, as it approaches 0.24, the model favors above maximal values of $\theta_{23}$, as is evident from Fig.~\ref{fig3}(f), (iv) in contrast to LMA phenomenology, the dark-LMA phenomenology of the model allows both CP conserving and CP violating parameter space as $J_{CP}$ can acquire both zero and non-zero values (Figs.~\ref{fig3}(c,d)). The results obtained for the effective Majorana mass parameter $m_{ee}$, as shown in Fig.~\ref{fig4}, are found to follow a trend similar to that observed for the LMA solutions and therefore lead to the same predictions.

\section{Conclusions}
We have successfully resolved the tension in the neutrino mixing angle $\theta_{12}$ that arises in the trimaximal mixing (TM$_2$) framework by incorporating minimal corrections from charged lepton sector. We have, also, identified the regions of parameter space consistent with the LMA and dark-LMA solutions. The phase $\delta$, arising from the charged lepton mixing matrix, provides an additional source of CP violation, alongside the neutrino sector. Consequently, the value of $|J_{CP}|$  can be as large as $0.13$. Using current oscillation and cosmological data on mixing parameters and sum of neutrino masses $\sum_i m_i$, we constrain the parameters $(\lambda,\delta)$ emanating from charged lepton corrections. For LMA with NH, there exist a upper bound at 3$\sigma$, $\lambda\le0.33$ and $\delta$ is constrained in the ranges $(20^\circ-90^\circ)\oplus (270^\circ-340^\circ)$. This upper bound is a consequence of the requirement that $\sin^2\theta_{23}$ must be greater than 0.43 (3$\sigma$ lower limit of $\sin^2\theta_{23}$\cite{Esteban:2024eli}) implying that the correction is dictated by current oscillation data. For dark-LMA, there exist a lower bound on $\lambda\ge0.24$ and phase $\delta$ constrained in the range $125^\circ<\delta<235^\circ$.  For both the LMA and dark-LMA solutions in the IH case, the effective mass parameter lies within the sensitivity range of current and future neutrinoless double beta decay experiments and can therefore be probed experimentally.

\section*{Acknowledgments} 
\noindent L. S. acknowledges the financial support provided by the Council of Scientific and Industrial Research (CSIR) vide letter No. 09/1196(18553)/2024-EMR-I. The authors, also, acknowledge the Department of Physics and Astronomical Science, Central University of Himachal Pradesh, for providing the necessary facilities to carry out this work.




\bibliographystyle{ieeetr}
\bibliography{ref}
\end{document}